\documentclass[useAMS,usenatbib,a4paper]{mn2e}

\usepackage{graphics}
\usepackage{hyperref}
\usepackage{graphicx}
\usepackage{times}
\usepackage{amsmath}
\usepackage{natbib}
\usepackage{wasysym}
\usepackage[usenames]{color}
\usepackage{epsfig}
\usepackage{epstopdf}
\usepackage{subfigure}
\usepackage{graphics}
\usepackage{longtable}
\usepackage{amssymb}
\usepackage{latexsym} 
\usepackage{wrapfig}
\usepackage{mathrsfs}
\usepackage{morefloats}
\usepackage{bm}
\usepackage[T1]{fontenc}
\usepackage{aecompl}
\usepackage{tikz}
\usetikzlibrary{positioning,shapes,fit,arrows}

\def\pupil{(0,0) ellipse (0.75cm and 1.5cm)}
\def\baselines{(3.0,0.0) ellipse (1cm and 1.8cm)}
\def\badphases{(3.1,-0.8) circle (0.6cm)}
\tikzset{main node/.style={circle,fill=cyan,draw,minimum size=0.75cm,inner sep=0pt}}

\title[Kernel Phase and Kernel Amplitude]
{Kernel Phase and Kernel Amplitude in Fizeau Imaging}
\author[B. J. S. Pope]
{Benjamin J. S. Pope\thanks{E-mail: benjamin.pope@physics.ox.ac.uk}\\
Oxford Astrophysics, Denys Wilkinson Building, Keble Rd, University of Oxford, Oxford OX1 3RH, UK}
\begin{document}
\maketitle

\begin{abstract}
Kernel phase interferometry is an approach to high angular resolution imaging which enhances the performance of speckle imaging with adaptive optics. Kernel phases are self-calibrating observables that generalize the idea of closure phases from non-redundant arrays to telescopes with arbitrarily shaped pupils, by considering a matrix-based approximation to the diffraction problem. In this paper I discuss the recent history of kernel phase, in particular in the matrix-based study of sparse arrays, and propose an analogous generalization of the closure amplitude to kernel amplitudes. This new approach can self-calibrate throughput and scintillation errors in optical imaging, which extends the power of kernel phase-like methods to symmetric targets where amplitude and not phase calibration can be a significant limitation, and will enable further developments in high angular resolution astronomy.
\end{abstract}


\section{Introduction}
\label{intro}

In imaging stars and their environments from the ground, under most circumstances the chief limitation on resolution and on sensitivity to faint structure is imposed by the turbulence of the Earth's atmosphere. This `seeing' introduces random delays in the phase of incoming light, so that when it arrives at a ground-based telescope, the wavefront is distorted and generates blurry, speckled images. Furthermore, as a higher-order effect, this wavefront distortion induces focussing and defocussing of the light in the Fresnel regime, so that there is also generally amplitude aberration, or scintillation, which is the reason that stars appear to twinkle. This is a major challenge in doing optical astronomy from the ground at high angular resolution, and in this paper I will review the history of techniques for ameliorating the effects of the atmosphere from speckle imaging; early hardware developments in non-redundant masking, where the telescope aperture is selectively masked out to facilitate optical calibration; and through to the more recent computer post-processing technique of kernel phase, where such a masking procedure is simulated purely with software. I will then present a generalization of the kernel phase idea, to kernel amplitudes, and discuss its applicability to present and future adaptive optics systems and space telescopes.

For mathematically modelling the atmospheric distortion of images, it is useful to consider any telescope, even one with a conventional filled aperture, as an interferometer, where light incident at different parts of the telescope is optically combined and caused to interfere. By measuring these interference fringes, the Van~Cittert-Zernike theorem \citep{1938Phy.....5..785Z} states that we can map out the Fourier transform of the intensity distribution of the source on the sky. This means that the Fourier plane is a natural representation for understanding optical imaging under most circumstances. Pairs of points in the telescope aperture are considered to form baselines, whose length and orientation determine the Fourier component to which the associated fringe is sensitive. Interferometers are classically considered to be either of the Michelson configuration, where the light is combined in the pupil plane, or the Fizeau configuration, where the light from all elements of the pupil is directly combined on a detector. 

In radio astronomy, it has long been possible to record the electric field received at each detector, and combine these pupil-plane signals in postprocessing via a correlator. In optical astronomy, because it is not possible to directly sample the waveform of the electric field as it is at radio frequencies, single-telescope filled-aperture imaging typically occurs in the Fizeau configuration. As a result, fringes from every pair of points in the pupil are superimposed to form the point spread function (PSF) of the telescope. This is true both of discrete sets of apertures, for example the double slit whose PSF is a sinusoidal fringe, and filled apertures, where the familiar circular telescope aperture's PSF is the Airy pattern. In the following, I will refer interchangeably to pupil elements and subapertures, and stations and antennae, using the appropriate terminology for specifically-optical applications and for historical radio applications respectively.

Sophisticated techniques for analysing images degraded by atmospheric turbulence have long been in use. By modelling the effects of the atmosphere on Fourier components of the image, speckle interferometry \citep{1970A&A.....6...85L} and speckle masking \citep{1977OptCo..21...55W} have been used to resolve systems at very high angular resolution while ameliorating the effects of the atmosphere. These techniques rely on the fact that, while the PSF is degraded in potentially complicated ways, the complex visibilities in its Fourier transform can be represented simply. To achieve this, one must be able to `freeze the seeing', i.e. to take exposures with appropriate signal-to-noise at timescales shorter than the characteristic timescale of the atmosphere's variations. 

A successful solution to the problem of seeing has been the technique of aperture masking, first used by \citet{fizeau1868}, where sections of the telescope aperture are deliberately blocked out to leave a non-redundant pattern of holes, i.e. a pattern of holes in which no two are separated by the same baseline, in order to make the resulting fringe pattern more resilient to aberrations and easier to calibrate. In a redundant aperture, on the other hand, multiple sets of subapertures generate the same baseline, and it is not so easy to disentangle their contributions to the resulting visibilities. Aperture masking has yielded some of the highest-resolution astronomical images obtained with a single-mirror telescope \citep{1999Natur.398..487T,1999ApJ...525L..97M}, relying on the idea of closure phase \citep{1958MNRAS.118..276J}, an interferometric self-calibration technique originating in radio astronomy. Often in interferometry, the dominant source of uncertainty is from phase aberrations in the pupil plane (i.e. for discrete interferometers, delay errors at individual stations), which are in general very large, originating from the ionosphere in the context of radio astronomy, or from atmospheric turbulence in the optical regime. These are especially important because, since the Fourier transform is Hermitian, the Fourier phases of a nonnegative real source (i.e. any astronomical image) encode only information about the asymmetric component of an image, and the moduli about its point-symmetric component \citep{2007NewAR..51..604M}. Therefore in the absence of prior information it is crucial to have access to both; regularly in astronomy, the issue is therefore to restore phase information which is typically more-degraded by the atmosphere. In this paper, we will discuss the context of methods for doing this, and then show that these can be generalized to self-calibrating both phase and amplitude information.

The key idea in interferometric self-calibration is the closure phase or (bispectral phase). These are sums of measured phases around closing triangles of baselines; because errors occur locally to each station, while astrophysical signals are encoded in correlations between stations, if you add the phases measured around such a triangle of baselines, the local phase errors cancel but the astrophysical signal adds. As a result, for a simple three-element interferometer, for the price of three phase observables corrupted by aberrations, one very stable observable can be obtained, which is often a significant advantage. This idea entered optical astronomy under the guise of `triple-correlation' speckle imaging \citep{1970A&A.....6...85L,1977OptCo..21...55W}, which was shown by \citet{1986OptCo..60..145R} to be the exact equivalent of the closure phase idea. While aperture masking in general requires, as with speckle imaging, that exposures are taken fast enough to freeze the seeing, adaptive optics can be used to dramatically increase the timescale of phase variations and effectively remove this limitation \citep{2006SPIE.6272E.103T}.

In radio astronomy, successive generations of calibration schemes have been able to use closure information in conjunction with the phases and amplitudes recorded at each antenna for self-calibration. \citet{2011A&A...527A.106S} classifies these into first-generation schemes using closure phases but not using individual receiver phases directly, second-generation schemes which iteratively self-calibrate the phase and gain at each receiver, and third-generation schemes which use the radio interferometry measurement equation to include direction-dependent effects. Because the phases and amplitudes of individual aperture elements cannot at present be easily and directly measured, optical imaging does not yet benefit from these second-~and third-generation calibration schemes. As a result, we are still limited to the classical closure relations in self-calibration, and I therefore seek to obtain the fullest and most detailed understanding of these quantities as is possible. 

Kernel phase interferometry was proposed for this reason by \citet{2010ApJ...724..464M}, who also coined the term itself. The key idea was a generalization of closure phase, using a matrix approach to show that a wider class of telescopes possess linearly self-calibrating quantities that can be used to improve the imaging performance of telescopes at resolutions close to the diffraction limit. Closure phases are shown to be a special case of this formalism, which in general applies to any standard Fizeau imaging system subject to only small aberrations. This technique has been used both for space telescopes such as the \emph{Hubble Space Telescope} \citep{2013ApJ...767..110P}, and also for ground-based adaptive optics systems \citep{2016MNRAS.455.1647P}.

In this paper, I discuss the closure and kernel quantities associated with general Fizeau interferometers and the relationship of previous methods to the more recent development of kernel phase interferometry. I propose the extension of this to kernel amplitudes, a natural generalization of the kernel phase theory which allows us to calibrate not only phase errors but also scintillation.

\section{Matrix Theory of Phase Self-Calibration}
\label{theory}

The \citet{1958MNRAS.118..276J} idea of closure phase rests on the fact that if two antennae,~1 and~2, would in the absence of noise record between them a complex visibility $C_{12}$ having phase $\Phi_{12}$, when they experience phase errors $\phi_1$ and $\phi_2$ then the recorded phase of the complex visibility between them is instead

\begin{equation}
\Phi'_{12} \equiv \Phi_{12} + \phi_1 - \phi_2.
\end{equation}

Therefore if we add the phases around a closing triangle of baselines, 

\begin{equation}
\Phi_{123} \equiv \Phi'_{12} + \Phi'_{23} + \Phi'_{31}
\end{equation}

\noindent contains each error $\phi_i$ exactly once positively and once negatively, and so the errors cancel out and the closure phase $\Phi_{123}$ is invariant. 

On the other hand, \citet{1991InvPr...7..261L} derives this closure phase relation in terms of a phase-transfer matrix, linearly mapping phase aberrations at discrete points in the pupil plane onto phases on the measured visibilities. This is a useful framework in which to consider generalizing the idea of closure phases to arbitrary pupils, which have kernel phases that work in the same way.

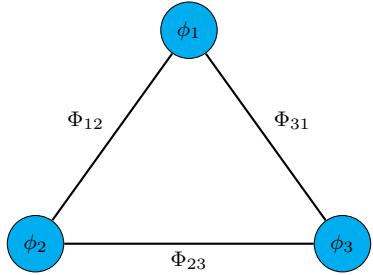
\begin{figure}
\centering
  \begin{tikzpicture}
    \node[main node] (1) {$\phi_1$};
    \node[main node] (2) [below left = 2.3cm and 1.5cm of 1]  {$\phi_2$};
    \node[main node] (3) [below right = 2.3cm and 1.5cm of 1] {$\phi_3$};

    \path[draw,thick]
    (1) edge node[above left] {$\Phi_{12}$} (2)
    (2) edge node[below] {$\Phi_{23}$} (3)
    (3) edge node[above right] {$\Phi_{31}$} (1);
\end{tikzpicture}
\caption{Lower-case $\phi_i$ are phase errors local to receiver $i$, while upper-case $\Phi_{ij}$ are the phases of the complex visibilities on baselines $ij$.}
\label{tels}
\end{figure}

As an example, consider a three-element array, with stations at the vertices of some triangle. The matrix

\begin{equation}
\label{transfermatrix}
\mathbf{A}_\phi \equiv \left(\begin{array}{ccc}
1&-1&0\\
0&1&-1\\
-1&0&1
\end{array}\right)
\end{equation}

\noindent encodes this linear relation between the three pupil samples and the three baselines that they generate: 

\begin{equation}
\left(\begin{array}{c}
\Phi_{12}^\prime\\ 
\Phi_{23}^\prime\\ 
\Phi_{31}^\prime
\end{array}\right) = \left(\begin{array}{ccc}
1&-1&0\\
0&1&-1\\
-1&0&1
\end{array}\right) \cdot \left(\begin{array}{c}
\phi_1\\ 
\phi_2\\ 
\phi_3
\end{array}\right) + \left(\begin{array}{c}
\Phi_{12}\\ 
\Phi_{23}\\ 
\Phi_{31}
\end{array}\right)
\end{equation}

\noindent where $\Phi_{ij}^\prime$ are the phases measured on the baseline $i,j$; $\phi_j$ is the phase aberration at station $j$; and $\Phi_{ij}$ the true phase as would be seen on an ideal unaberrated interferometer on baseline $ij$. This is illustrated in Figure~\ref{tels}.

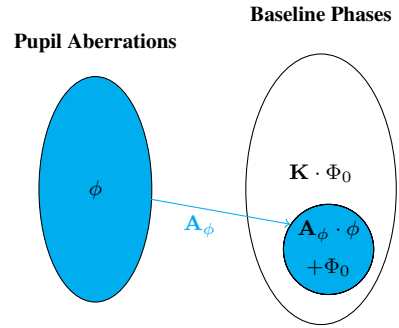
\begin{figure}
\centering
    \begin{tikzpicture}
      \begin{scope}
    \fill[cyan] \pupil;
      \end{scope}
      \begin{scope}
    \fill[cyan,minimum size=25cm] \badphases;
      \end{scope}

    \draw \pupil node[text=black] (pupil) {$\phi$}
          \baselines node[text=black,above] (base) {$\mathbf{K}\cdot\Phi_0$} 
          \badphases node[text=black,above] (im) {$\mathbf{A_\phi}\cdot\phi$}
          \badphases node[below=0.02cm,text=black] (imb) {$ + \Phi_0$} (im);
    \draw[->,cyan] (pupil) -- node[pos=0.5,below] {$\mathbf{A}_\phi$} (im) ;
    \node[above=1.7cm of base,font=\color{black}\bfseries] {Baseline Phases};
    \node[above=1.5cm of pupil,font=\color{black}\bfseries] {Pupil Aberrations};
    \end{tikzpicture}
\caption{Relations between spaces of phases on pupil and baselines. The operator $\mathbf{A}_\phi$ injects phase aberrations $\phi$ into a subspace of baseline phases $\Phi$. There is another subspace of baseline phases, $\mathbf{K}_\phi \cdot \Phi_0$, which are not affected by these phase aberrations, and we call these kernel phases.}
\label{diagram}
\end{figure}

The sum of each column of this matrix is zero, so that the row-vector

\begin{equation}
C_\phi \equiv \left(\begin{array}{ccc}
1&1&1
\end{array}\right)
\label{cphi}
\end{equation}

\noindent must annihilate $\mathbf{A}_\phi$, i.e., $C_\phi \cdot \mathbf{A}_\phi = \mathbf{0}$. An operator $C_\phi$ satisfying this condition is called a `left-kernel' operator for $\mathbf{A}_\phi$, whose rows span a `left null space' or `cokernel space' for $\mathbf{A}_\phi$, which motivates the use of the term `kernel phase' for quantities of the form $C_\phi \cdot \Phi$\footnote{We need to choose a left-kernel as $\mathbf{A}_\phi$ and $\mathbf{A}_\phi^{T}$ map vectors between different spaces and multiplication is not commutative.}, i.e. for linear combinations of baseline phases $\Phi$ that `live in' the cokernel of $\mathbf{A}_\phi$. This will not in general be a row-vector for interferometric arrays with more elements, but a matrix whose rows consist of vectors like in Equation~\ref{cphi} which encode closure relations. A diagram illustrating the relations of these spaces is provided for clarity in Figure~\ref{diagram}.

By summing phases around a closed triangle as above, phase aberrations do not propagate to the new closure phase observable, which accordingly is much more reliable than raw phases and can be used to anchor image reconstruction and model-fitting. In general, we will consider pupil phases as being relative to the phase at an arbitrarily chosen point, and thereby eliminate this arbitrary degree of freedom, so that pupil sample~1 is taken to have identically zero phase, and the matrix in Equation~\ref{transfermatrix} becomes 

\begin{equation}
\label{transfermatrix2}
\mathbf{A}_\phi \equiv \left(\begin{array}{cc}
-1&0\\
1&-1\\
0&1
\end{array}\right)
\end{equation}

\noindent which is annihilated by $C_\phi$ in the same way as before. 

Closure phases do not necessarily have to be taken around a triangle: \citet{1991InvPr...7..261L} identifies this cokernel of $\mathbf{A}_\phi$ as the definition of the closure operator, rather than as an alternative derivation. This operator then generates all orders of closure phase, not just the set of closing triangles, including for example loops containing four or more elements. The kernel of $\mathbf{A}_\phi$ can be found numerically by singular value decomposition (SVD), by arranging as a matrix the singular vectors whose singular values are zero. For a non-redundant aperture, this left-kernel operator's rows consist of a basis of vectors spanning the space of closure (kernel) phases.

In order for this self-calibration to be exact, closure phases must be taken around closing sets of baselines whose phases map linearly onto the measured baseline phases, which is true for non-redundant baselines, where the two stations are not separated by the same vector as any other pair of holes (or for completeness, also in the special case of phasors which have identical amplitudes on doubly-redundant baselines). The phase noise introduced to closure phases by aberrations in a circular redundant aperture scale as $\sim (D/{r_{coh}})$, where $D$ is the telescope diameter and ${r_{coh}}$ the atmospheric coherence scale \citep{1988AJ.....95.1278R}. Avoiding this redundancy noise has traditionally necessitated the use of a non-redundant mask, but the advent of modern adaptive optics means that phase-based interferometric techniques can now be extended at high fidelity to filled apertures of high redundancy.

The idea of kernel phase interferometry is that, while these relations are no longer exact for redundant arrays, they approximately hold true for small phase aberrations \citep{2010ApJ...724..464M}. This now requires that each baseline be weighted by a diagonal matrix $\mathbf{R}$, whose entries are the redundancies of each baseline (i.e. how many pairs of subapertures correspond to that baseline), such that now

\begin{equation}
\Phi' \approx \mathbf{R}^{-1} \cdot \mathbf{A}_\phi \cdot \phi + \Phi
\end{equation}

\noindent and the kernel matrix $\mathbf{K}_\phi$ is found by SVD of $\mathbf{R}^{-1} \cdot \mathbf{A}_\phi$. As in this general case $\mathbf{A}_\phi$ never appears without $\mathbf{R}^{-1}$, it is standard to redefine the transfer matrix $\mathbf{A}_\phi$ as the originally-defined matrix of ones and zeros, rescaled by $\mathbf{R}^{-1}$. This is the redundant, weighted equivalent of the phase aberration matrix in \citet{1991InvPr...7..261L}, so that the kernel phases are a generalization of the higher-order closure phases of a filled pupil. We will use this rescaling in the remainder of this text.

Such a kernel space necessarily exists for any linear imaging system with more baselines than pupil sample points, according to the fundamental theorem of linear algebra: aberrations from~$p$~pupil sample points can map onto at most a space of dimension $p$ over the baselines, which will ordinarily have a much larger dimension than the pupil samples. This implies that we should look for further generalizations to other observables and imaging systems.

\section{Kernel Amplitudes}
\label{kernel_amplitudes}

In this Section, we will show that there is an analogous quantity to the kernel phase in the case of amplitudes, which we will call the kernel amplitude. This allows us to self-calibrate both phase and amplitude information in arbitrary Fizeau interferometers.

\citet{1960Obs....80..153T} first introduced the concept of a four-element interferometer closure amplitude, which was subsequently systematically studied and described by \citet{1980Natur.285..137R}. This quantity is an extension of the idea of closure phase to visibilities which are affected by uncertainties in the gain on individual antennae, or equivalently the throughput and reflectivity of different optical paths in optical interferometry. In the following, for ease of reference, `visibility' is used to mean the modulus of the complex visibility unless stated otherwise, and similarly, `gain' for the modulus of the complex gain. The derivation is similar to the \citet{1958MNRAS.118..276J} idea described at the beginning of Section~\ref{theory}, except that for each visibility $V_{ij}$ between telescopes $i$~and~$j$, there are gain errors $g_i$ and $g_j$ which multiply rather than add, so that the measured visibility is 

\begin{equation}
\label{vismeas}
V_{ij}' = V_{ij}\cdot g_i \cdot g_j
\end{equation}

\noindent Therefore the quantity 

\begin{equation}
\label{standard_closure_vis}
V_{1234} \equiv \dfrac{V_{12}' V_{34}'}{V_{13}' V_{24}'}
\end{equation}

\noindent multiplies each gain error through once on the numerator and once on the denominator; these cancel out and the closure amplitude $V_{1234}$ is seen to be invariant with respect to this type of error.

The same matrix approach as for closure phases can be used to derive closure amplitudes. It is useful to deal with log-visibilities on each baseline, and log-gains on antennae, so that these quantities add and subtract rather than multiply and divide. \citet{1991InvPr...7..261L} notes that while the phase error on a baseline is the difference in phase between the stations, the measured log-visibility of a point source is the sum of the log-gains on each telescope. Therefore the elements of the transfer matrix $\mathbf{A}_V$ are just the absolute values of the corresponding elements of $\mathbf{A}_\phi$. 

The linear gain transfer operator $\mathbf{A}_V$ for a non-redundant four-element interferometer is therefore

\begin{equation}
\mathbf{A}_V \equiv \left(\begin{array}{cccc}
1&1&0&0\\
0&1&1&0\\
0&0&1&1\\
1&0&0&1
\end{array}\right)
\end{equation}

\noindent with a transfer equation

\begin{align}
\label{transfer}
\left(\begin{array}{c}
\log(V_{12}^\prime)\\ 
\log(V_{23}^\prime)\\ 
\log(V_{34}^\prime)\\
\log(V_{41}^\prime)
\end{array}\right) &= \left(\begin{array}{cccc}
1&1&0&0\\
0&1&1&0\\
0&0&1&1\\
1&0&0&1
\end{array}\right) \cdot \left(\begin{array}{c}
\log(g_1)\\ 
\log(g_2)\\ 
\log(g_3)\\
\log(g_4)
\end{array}\right) \\&+ \left(\begin{array}{c}
\log(V_{12})\\ 
\log(V_{23})\\ 
\log(V_{34})\\
\log(V_{41})
\end{array}\right)
\end{align}

\noindent which we write in vector notation as 

\begin{equation}
\mathbf{V}' = \mathbf{A}_V \cdot \mathbf{\log(g)} + \mathbf{V}
\end{equation}

\noindent and we see that the operator

\begin{equation}
C_V \equiv \left(\begin{array}{cccc}
1&-1&1&-1
\end{array}\right)
\end{equation}

\noindent annihilates $\mathbf{A}_V$, so that

\begin{align}
\log(V_{1234}) &\equiv C_V \cdot \mathbf{V'} \\
&= \log(V_{12}) + \log(V_{34}) - \log(V_{13}) - \log(V_{24})
\end{align}

\noindent which is the standard four-element closure amplitude relation from Equation~\ref{standard_closure_vis}. Gains can be chosen to be relative to a fixed gain of unity as was done with phases in Section~\ref{theory} without loss of generality, and are shown above in full unreduced form above for clarity. In software implementations, I will follow the kernel phase example and fix the gain of aperture~1 at unity. 

This formalism applies to only single baselines: the visibility of a redundant baseline is the modulus of the sum of the individual phasors, so that we want to add the visibilities and not the log-visibilities. As a result, the non-redundant calculation of \citet{1991InvPr...7..261L} does not hold in the redundant case, and we must consider differential quantities in order to linearize the problem. First we note the Taylor expansion

\begin{equation}
\label{linear_approx}
\log(g) \approx \log(1+\Delta v) \approx \Delta v
\end{equation}

\noindent which means that gain perturbations away from unity map approximately linearly onto the corresponding, generally-redundant baselines. We are therefore able to consider the linear gain transfer equation

\begin{equation}
\Delta V' = \mathbf{R}^{-1}\cdot \mathbf{A}_V \cdot {\Delta v} + \Delta V
\end{equation}

\noindent which can be considered as the generalization of the \citet{2010ApJ...724..464M}-style approximation to the \citet{1991InvPr...7..261L} matrix-based closure amplitudes. We therefore analogously construct $\mathbf{K}_V$ by SVD such that $\mathbf{K}_V \cdot (\mathbf{R}^{-1}\cdot \mathbf{A}_V) = 0$, and take this to be the kernel amplitude operator.

By the rank-nullity theorem, there will be at least $n_{b} - n_{p}$ kernel amplitudes, where $n_{b}$ is the number of baselines and $n_{p}$ the number of pupil samples. This is in general large: in the non-redundant case, $n_{b}$ is combinatorically larger than $n_{p}$, and while this is lower for redundant cases, depending on the detailed pupil geometry we may recover as much as $90\%$ of the amplitude information as is done with kernel phases \citep{2013aoel.confE...6M}. 

The same transfer matrix formalism used to derive kernel phases can also be inverted for wavefront sensing \citep{2013PASP..125..422M,2014MNRAS.440..125P}. In this framework, a Moore-Penrose pseudoinverse is generated by SVD of the transfer matrix, and used to map measured $u, v$ phases back onto the pupil sampling points. This requires not only that the kernel phase approximation holds, but also that the pupil itself is not inversion-symmetric; otherwise, it is only possible to sense aberration modes that are odd under inversion, and even modes are unsensed. This generalizes to the kernel amplitude formalism, although as with the case of phases, there are detailed issues surrounding matrix stability and convergence that mean that a proper examination must be left to future studies.

As the quantities we are considering in the redundant case are differential amplitudes (i.e. $\Delta V$ relative to a standard visibility), it is necessary to have some standard visibility model against which these differences can be taken. There are several possible ways to do this: most simply, an image of a fiducial point source can be used as a calibrator, such that the differential visibilities we choose when making the kernel amplitudes are the differences between the science target and the calibrator. In this approach, we calibrate pupil-dependent aberrations, but baseline-dependent aberrations are not corrected, and we expect to have potentially substantial residual systematic errors.

A sensible and self-consistent alternative to this is to use the redundancy information already encoded in the matrix $\mathbf{R}$. For a redundant pupil consisting of discrete elements, the normalized visibility of a point source is given by the magnitude of the optical transfer function (OTF) on any baseline, which is proportional to the redundancy on that baseline and normalized to unity at the origin. Thus the pupil model itself generates a starting point for calculating differential visibilities. In this case, we can then use the kernel amplitudes measured on a known point source to calibrate baseline-dependent systematics that would otherwise not be corrected.

A treatment of the full general case of phase and amplitude errors in arbitrary linear imaging systems is beyond the scope of this paper, but we may start to sketch out a roadmap to finding such a theory. If we write the total aberrations as a concatenation of phase and amplitude error vectors, i.e.

\begin{equation}
\mathbf{a}^T \equiv [\mathbf{\phi}^T, \mathbf{\Delta V}^T]
\end{equation}

then the full transfer matrix for a linear Fizeau interferometer is the block diagonal matrix 

\begin{equation}
\mathbf{A} \equiv \left(\begin{array}{cc}
\mathbf{A}_{\phi}&\mathbf{0}\\
\mathbf{0}&\mathbf{A}_{V}
\end{array}\right).
\end{equation}

The kernel operator for this matrix is also a block-diagonal matrix of the kernel operators of each block separately, so that at linear order the full kernel phase and amplitude problem is itself separable. For systems which mix phase and amplitude errors, the above matrix will instead have non-zero off-diagonal elements, but will have a set of kernel phase-amplitude quantities which nevertheless permit self-calibration.

\section{Phase Noise and Kernel Amplitudes}
\label{phasenoise}

In the above analysis, I have noted that the effects of phase errors on visibilities enter at second order. Given that phase noise generally dominates over amplitude noise in ground-based observations, it is therefore important to examine these effects at higher order, and the possibilities of linear-order phase effects, which have not been hitherto included in the kernel amplitude derivation.

\begin{figure}
\center
\includegraphics[width=0.5\textwidth]{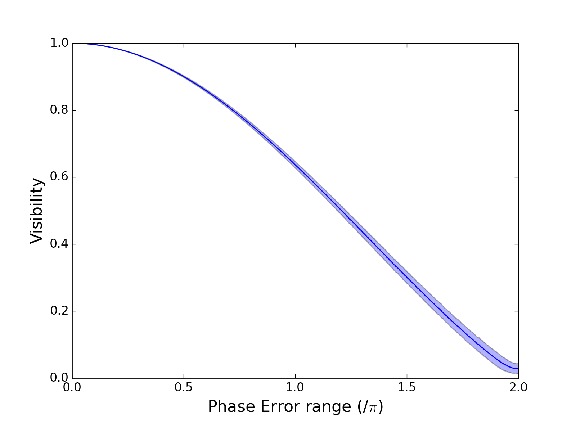}
\caption{Visibility bias, with uncertainties ($1\sigma$), for a 1000-redundancy baseline, with equal visibilities and phases drawn from a uniform distribution of width $s$ on the horizontal axis.}
\label{visbias}
\end{figure}

The principal effect of phase aberrations on visibilities is to bias them downwards: because of the triangle inequality, the length of the sum of $N$ vectors with any orientations is necessarily smaller than when they are all lined up. In Figure~\ref{visbias}, I show the results of a simple simulation to illustrate this. We add 1000 vectors (i.e., simulate a redundancy-1000 baseline) with the same length, and phases drawn from a uniform distribution of width $s$, which is displayed on the horizontal axis. We see that the visibility bias is smooth, with a small uncertainty at each scale, growing stronger and more uncertain at higher phase noise levels. Importantly, near zero phase error, the function is locally flat, and so the block-diagonal transfer matrix approximation holds. 

\begin{figure}
\center
\includegraphics[width=0.5\textwidth]{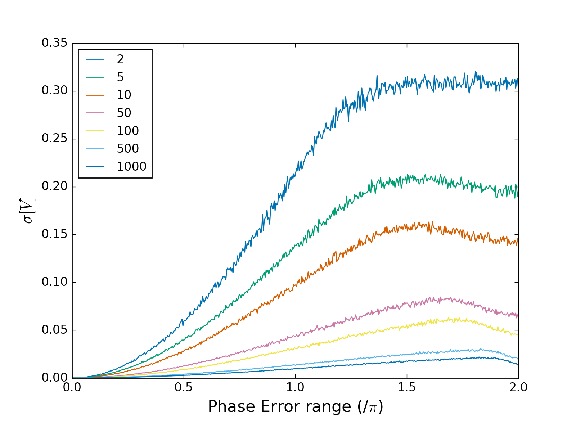}
\caption{Visibility bias error: standard deviations of visibilities across 5000 simulations on a single baseline, for a range of baseline redundancies (listed in legend).}
\label{visbiaserr}
\end{figure}

We see that at higher levels of phase noise ($\sim 0.5$~rad), the bias increases, as does its variance. We see this error plotted for a range of redundancies in Figure~\ref{visbiaserr}. The error is much worse for lower redundancies than higher redundancies: as we add more vectors, the central limit theorem constrains the variance of the outcome more strongly. 

In cases such as these, it seems likely that the moderate phase-to-amplitude off-diagonal block may be numerically simulated for a given visibility bias, which must be measured in practice from observations and cannot be determined a priori like the standard phase and visibility transfer matrices. The singular vectors of this matrix will generate a kernel \emph{phase and amplitude} basis, though this added layer of complexity may not be of practical utility.

\section{Simulation}
\label{sims}

In the following, I present the results of a number of simulations to determine the performance of kernel amplitudes relative to raw visibilities. As diffraction simulations to generate PSFs are much more computationally expensive than subsequently convolving these PSFs with binary or disk templates, I adopt the approach of using a single ensemble of PSFs and introducing successively fainter versions of a binary or disk, with all other parameters held fixed. As a result, the simulations in this section are intended to display the relative performance of the two techniques under identical conditions, rather than to forecast actual contrasts achievable in real observations. In Section~\ref{binaries} we search for a single point-like companion; in Section~\ref{phasenoisesims}, we do the same as before, but testing for the effects of phase noise on kernel amplitudes; and in Section~\ref{disks}, we illustrate how kernel amplitudes are sensitive to a symmetric target (which kernel phases alone are necessarily insensitive to), in this case a circumstellar debris disk. In many but not cases the kernel amplitude approach produces a less biased estimate, with more conservative errorbars, and typically extends comparable sensitivity to fainter contrasts than using raw visibilities alone.  

\begin{figure}
\center
\includegraphics[width=0.5\textwidth]{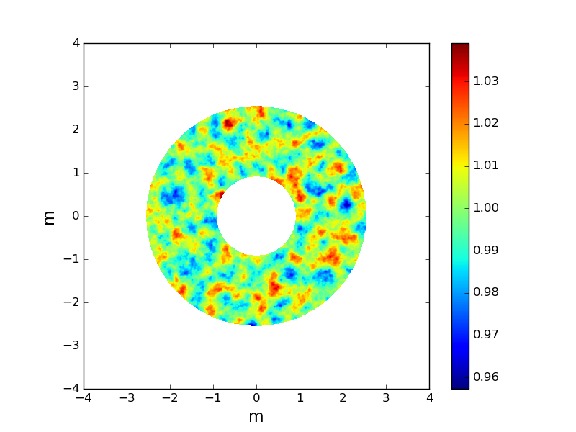}
\caption{Example amplitude screen. Colourmap in dimensionless units of relative amplitude.}
\label{ampscreen}
\end{figure}

In Sections~\ref{binaries} to~\ref{disks}, I have generated a sample of 100 amplitude screens, calculated using the analytic \citet{1975RaSc...10..155R} power spectrum for Fresnel diffraction from a single layer of turbulence,

\begin{equation}
\label{fresnel}
P_S = 4 P_\phi \sin^2 (\kappa_r^2 / \kappa_F^2) 
\end{equation}

\noindent where $P_\phi$ is drawn from a Kolmogorov spectrum with $r_0 = 1\text{mm}$, $\kappa_r$ is an isotropic spatial frequency and $\kappa_F$ the Fresnel wavenumber $\kappa_F = \sqrt(4\pi/\lambda z)$, and $z$ is the height of the turbulent layer, which we choose to be 30~km. This is simulated across a pupil of the same dimensions as the Palomar Hale 200-Inch Telescope (outer diameter 5.093~m, secondary diameter 1.829~m), and then scaled so that the amplitude is positive-definite with variations of a specified scale. These parameters are generous but not unreasonable, but specifically chosen to guarantee sufficiently low spatial frequency pupil plane speckles that these are computationally efficient to sample. A more detailed simulation would require a much denser discrete pupil model and correspondingly larger interferometric field of view. I also note that while these speckles are rather large for the fiducial Hale 200-Inch telescope model used here, a smaller telescope will more often experience scintillation speckles of a similar scale relative to the telescope diameter. An example of such an amplitude screen is shown in Figure~\ref{ampscreen}. 

These screens then multiply the pupil binary mask function to generate the input pupil plane. This is Fourier transformed to generate PSFs, which are then convolved with source functions to generate images. No shot or detector readout noise is added. I then use \textsc{MultiNest} \citep{2009MNRAS.398.1601F,2013arXiv1306.2144F} to conduct a Bayesian fit of an analytic parametric models to the ensemble mean of the squared visibilities or kernel amplitudes extracted from these. 

\subsection{Binaries}
\label{binaries}

\begin{figure*}
\centering
\includegraphics[width=\linewidth]{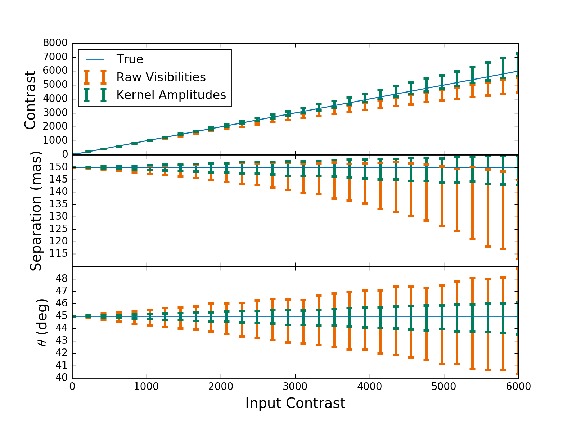}
\caption{Kernel amplitude (dark teal) and raw visibility (bright orange) best fits to the ensemble mean of 50 sets as a function of input binary contrast, under conditions of $10\%$ scintillation and no phase noise. Both methods obtain precise estimates at low contrasts, but at higher contrasts the estimate from fitting to raw visibilities is increasingly biased, while the kernel amplitudes maintain a close 1:1 correlation.}
\label{sims_all}
\end{figure*}

To test the stability of kernel amplitudes, we first simulate binary systems without any phase aberrations, from Kolmogorov spectrum scintillation screens of peak-to-trough variation~$10\%$ relative to the unaberrated image, to explore the effect of scintillation directly. (This is a high but not very high value, with typical scintillation ranging up to $\sim 1\%$ in RMS photometric amplitude with a more complicated spatial distribution). In the following, PSFs generated as above are shifted and added to simulate a binary at a varying contrast and a separation of 150~mas ($\sim 1.4 \lambda/D$). I then fit an analytic binary model to the ensemble mean of kernel amplitudes or squared visibilities from these, taking a uniform prior in contrast (the ratio of primary to secondary flux) between $c/2$ and $2c$ for each input contrast $c$. The results are displayed in Figure~\ref{sims_all}.

I note that the fits from mean kernel amplitudes have smaller errorbars than those from raw visibilities, and that in all cases these overlap with the true value. This is not the case for visibilities: these fits are perturbed by systematics that are not well-accounted for and retrieve larger errorbars in position and a downward bias in contrast increasing as we push to the fainter end.

I expect that this effect is more pronounced at lower contrasts for higher-amplitude scintillations, and vice versa, and illustrates that while under good observing conditions visibilities behave well, that kernel amplitudes allow us to push to greater precision and sensitivity at a given level of scintillation. 

\subsection{Phase Noise}
\label{phasenoisesims}

To determine the effect of phase noise on the practical use of kernel amplitudes, I re-did the binary simulations in Section~\ref{binaries} with the same parameters, except with the pupil screen changed to a 1.5~rad peak-to-peak variation, where these are drawn from the same Gaussian random fields as in Section~\ref{binaries} scaled by a square root factor for consistency (as atmospheric scintillation ordinarily enters at second and higher order in the phase variation). These simulations include no amplitude noise. Binary models are then fitted as above, using the same kernel amplitude operator.

\begin{figure*}
\centering
\includegraphics[width=\linewidth]{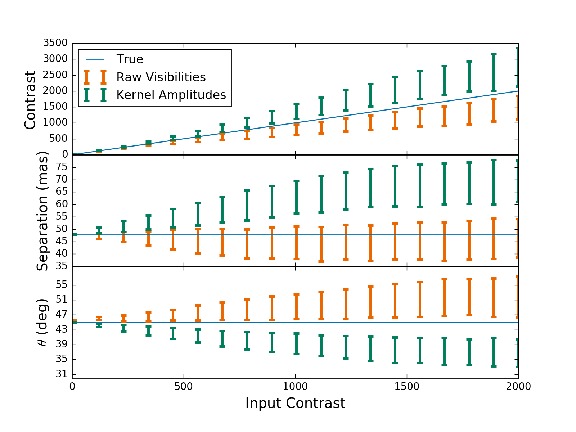}
\caption{Ensemble mean kernel amplitude (dark teal) and raw visibility (bright orange) best fits as a function of input binary contrast, under conditions of phase noise with no amplitude aberrations.}
\label{sims_phase}
\end{figure*}

As seen in Figure~\ref{sims_phase}, posterior parameter uncertainties are successively degraded to higher contrasts for both raw visibilities and kernel amplitudes, with no obvious indication of severe failure on either part out to high contrast and with neither obviously performing better or worse than the other in either contrast or angle, although visibilities perform better at reconstructing the separation. From these results, I take that the kernel amplitude is not much more adversely affected than the raw visibility, which is in line with our expectations that it should have no special self-calibrating property in this regard.

\subsection{Disks}
\label{disks}

In addition to simulating binaries, which are inherently point-asymmetric images, I also simulate elliptical annuli as an example of centrosymmetric circumstellar disks. PSFs are generated as in Section~\ref{binaries}, and convolved with a top-hat-like model which takes the value unity between two similar ellipses, and zero everywhere else. These are scaled by the desired contrast (taken as the ratio of total unresolved flux to disk flux) and added to the original PSF. An elliptical annulus has the analytic visibility function of the normalized difference between two ellipses, with five parameters: semi-major axis, eccentricity, position angle of the major axis, contrast ratio, and thickness (ratio of the difference in semi-major axes of the inner and outer ellipse, normalized to the outer).

\begin{figure*}
\centering
\includegraphics[width=\linewidth]{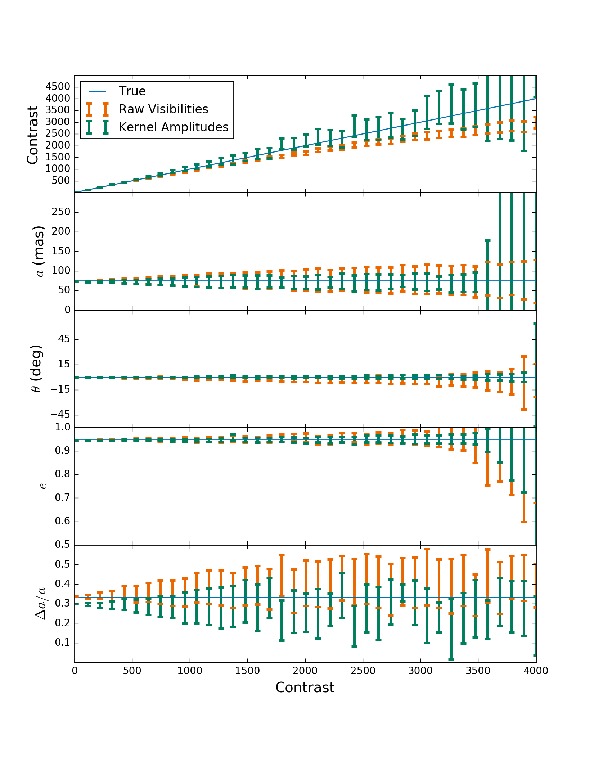}
\caption{Ensemble mean kernel amplitude (dark teal) and raw visibility (bright orange) best fits as a function of input binary contrast.}
\label{sims_disk}
\end{figure*}

We fit these in the same way as with the binaries in Section~\ref{binaries}, except that with five parameters it is no longer feasible to fit each realization separately, so I only fit to the mean signal at each contrast. As seen in Figure~\ref{sims_disk}, both kernel amplitudes and raw visibilities perform well at low contrasts, and then gradually lose effectiveness at higher contrasts. We see that raw visibility fits are biased low in contrast, with many-$\sigma$ systematic discrepancies from the true contrasts, while for kernel amplitudes, error bars get larger but always include the true value. The two methods seem to perform similarly well as a function of contrast for $\Delta a/a$, with visibilities biased a little high and kernel amplitudes low, while for position angle $\theta$, eccentricity $e$, and semi-major axis $a$ (except at high contrast), kernel amplitudes offer more precise and accurate measurements out to high contrast.

As with the binary case in Section~\ref{binaries}, the amplitude errors introduced are in practice very large in an observational context, to illustrate the effect, and higher contrasts will be reachable with correspondingly lower amplitude noise. It is not apparent from Figure~\ref{sims_disk} that kernel amplitudes reach a dramatically better performance than raw visibilities for disks, but they nevertheless appear to be somewhat more accurate, and often more precise, than using raw visibilities. 

\section{Discussion}
\label{discussion}

Kernel amplitudes have promise for improving observations in optical regimes where small-to-moderate amplitude aberrations are a limiting noise source. While phase aberrations are under normal circumstances a much more severe problem, correcting amplitudes is useful under several circumstances:\\

a) Amplitude errors arising from plane-wave atmospheric scintillation impose limitations on the performance of adaptive optics \citep{1994Natur.368..203A,2006JOSAA..23.2602L}. In cases where these cannot be otherwise corrected, imaging performance may be enhanced by anchoring models or image reconstructions with kernel phases and amplitudes.\\

b) In certain cases it is important to establish with great precision the optical visibilities of an astrophysical object whose angular size is close to the diffraction limit of available telescopes. For instance, in studying dusty circumstellar environments with aperture masking, it is useful to consider differential visibilities between two polarization channels \citep{2012Natur.484..220N, 2015MNRAS.447.2894N}. In such situations, uncertainties in the throughput of each optical channel contribute significantly to the error budget. While future instruments may be able to directly measure this with pupil remapping \citep{2006MNRAS.373..747P}, for the present and near-term a software solution may be preferable. Differential (polarimetric) kernel amplitudes permit the extension of the experiments described above to, for example, the SPHERE-ZIMPOL extreme-AO polarimetric imager \citep{2014SPIE.9147E..1RL}. By using a filled pupil and extreme adaptive optics, kernel phases already enable us to push towards fainter objects than are accessible with the large throughput losses from aperture masking on comparable instruments. We have shown kernel amplitudes have the potential to enhance this effect in situations where scintillation is also a problem.

\section{Conclusion}
\label{conclusion}

With kernel amplitudes and kernel phases, we now have a full formalism to describe all self-calibrating observables in a direct Fizeau imaging system, such as most full-aperture telescopes. This means that both even and odd symmetry components of an image reconstruction can now be directly based upon self-calibrating observations. By including both even and odd image components, we will be able to improve existing methods of imaging stars, circumstellar and protoplanetary disks, and other astrophysical sources with both extended symmetric structure and embedded inhomogeneities. This may be of limited use in standard imaging, but in specialist cases where scintillation is the limiting factor (such as in Section~\ref{discussion}) this may be a step forward. I hope that this will prove useful in forthcoming imaging campaigns with adaptive optics and space telescopes. 

The fundamental theorem of linear algebra also implies that similar self-calibrating observables may be discoverable for other linear imaging systems, whenever the relevant measurement basis (e.g. baselines) is of substantially higher dimension than the pupil samples which generate it. It will be of compelling future interest to establish the extent to which this is true of coronagraphic systems, whose sensitivity to small phase and amplitude aberrations is a key limiting factor in the search for exoplanets. 

In the interest of open science, \textsc{IPython} Notebooks and \textsc{Python} scripts implementing the simulations used in this paper are available at \url{https://github.com/benjaminpope/pysco}, under a GNU General Public License (v3).

\section*{Acknowledgements}
\label{acknowledgements}
I am grateful to Laurent Pueyo, Frantz Martinache, Peter Tuthill, Anand Sivaramakrishnan, Colin Norman, James Osborn, James Lloyd, John Monnier, Suzanne Aigrain, and Pat Roche for their helpful comments. In addition to this, I owe thanks to the reviewer Chris Haniff for his thorough work. I would also like to thank Balliol College and the Clarendon Fund for their financial support.

This research has made use of the \textsc{IPython} package \citep{PER_GRA:2007}, and \textsc{matplotlib}, a \textsc{Python} library for publication quality graphics \citep{Hunter:2007}.

\bibliographystyle{aa}
\bibliography{ms}

\end{document}